\documentclass{aa}
\usepackage{graphicx}
\begin{document}

\newcommand{\apjl}{ApJ Let.}
\newcommand{\apj}{ApJ}
\newcommand{\mnras}{MNRAS}
\newcommand{\nat}{Nature}
\newcommand{\apjs}{ApJ Supp.}
\newcommand{\aap}{A\&A}
\newcommand{\bain}{Bull. Astr. Soc. Netherlands}
\newcommand{\de}{\delta}
\newcommand{\D}{\Delta}
\newcommand{\p}{\partial}
\newcommand{\La}{\Lambda}
\newcommand{\up}{\Upsilon}
\newcommand{\Om}{\Omega}
\newcommand{\be}{\begin{equation}}
\newcommand{\ee}{\end{equation}}
\newcommand{\al}{\alpha}
\newcommand{\ka}{\kappa}
\newcommand{\g}{\gamma}
\newcommand{\om}{\omega}
\newcommand{\ep}{\epsilon}
\newcommand{\va}{\varepsilon}
\newcommand{\prd}{phys. Rev. D}

\title{Lower limits on the maximum orbital frequency around rotating 
strange stars}


\author{Dorota Gondek-Rosi{\'n}ska\inst{1,2}
\and Nikolaos Stergioulas\inst{3}
\and Tomasz Bulik\inst{2}
\and W\l odek Klu\'zniak\inst{4}
\and Eric Gourgoulhon\inst{1}}

%

\institute{D{\'e}partement d'Astrophysique Relativiste et de Cosmologie 
 UMR 8629 du CNRS, Observatoire de Paris, F-92195 Meudon  Cedex, France
\and Nicolaus Copernicus Astronomical Center, Bartycka 18, 00-716 Warszawa,
 Poland
\and
Department of Physics, Aristotle University of Thessaloniki,
 54006 Thessaloniki, Greece
\and
Institute of Astronomy, Zielona G\'ora University, ul. Lubuska 2, 
PL-65265 Zielona G{\'o}ra, Poland}
\offprints{D. Gondek-Rosi{\'nska}, \email{Dorota.Gondek@obspm.fr}}

\date{Received, Accepted}

\abstract{ Observations of kHz quasi-periodic oscillations  (QPOs) in
the X-ray fluxes of low-mass X-ray binaries (LMXBs) have been used in
attempts to constrain the external metric of the
compact members of these binaries, as well as their masses
and the equation of state of matter at supranuclear denisties.
 We compute the maximum orbital
frequency of stable circular motion around uniformly rotating strange
stars described by the MIT bag model. 
 The calculations are performed for both
normal and supramassive constant baryon mass sequences of strange
stars rotating at all possible rates.  We find the lower limits on the
maximum orbital frequency and discuss them for a range of masses and
for all rotational frequencies allowed in the model considered. 
We show that for  slowly and
moderately rotating strange stars the maximum value of orbital
frequency can be a good indicator of the mass of the compact object.
However, for rapidly rotating strange stars the same value of orbital
frequency in the innermost stable circular orbit is obtained for
stars with  masses ranging from that of a planetoid to about three
solar masses. At sufficiently high rotation rates of the strange star, 
the rotational period alone constrains the stellar mass to a
surprisingly narrow range.
 }

\maketitle

\section{Introduction}
The recently discovered kHz quasi periodic oscillations may be used
as a probe of the inner regions of accretion disks in compact stars
and hence also of some of the properties of the central object.  For
example, these oscillations have been used to derive estimates of the
mass of the central source in some sources (e.g., Kaaret et al., 1997;
Zhang et al., 1998; Klu{\'z}niak 1998). All these estimates assume that
the maximum observed QPO frequency is the orbital frequency at the
innermost stable circular orbit (ISCO), as suggested by Klu{\'z}niak
et al. (1990).

In the first study of the possible presence of strange stars
in those LMXBs which exhibit kHz QPOs, Bulik et al. (1999a,b) 
showed that slowly rotating strange
stars, described by the simple MIT bag model with massless and
non-interacting quarks, have orbital frequencies at the marginally
stable orbit higher than the maximum frequency of 1.07  kHz in
4U 1820-30, reported by Zhang et al. (1998).  However, the ISCO
frequencies can be as low as 1 kHz
 when more sophisticated models of quark matter
(MIT bag with massive
strange quarks and lowest order QCD interactions) and/or rapid
stellar rotation are taken into account 
(Stergioulas et al., 1999; Zdunik et al., 2000a,b). The main result of these
considerations was that the lowest orbital frequency at the ISCO is
attained either for non-rotating massive configurations 
close to their  maximum mass limit, or for
configurations at the Keplerian mass-shedding limit. In order to obtain
an  orbital frequency in the ISCO as low as 1.07 kHz for slowly rotating
strange stars one has to consider a specific set of (rather extreme)
MIT bag model parameters (Zdunik et al. 2000a, b) allowing
gravitational masses higher than 2.2 $M_\odot$.  In contrast, for
stars rotating close to the Keplerian limit, a low orbital frequency
at the ISCO can be obtained for a broad range of stellar masses
(Stergioulas et al., 1999).

A new description of quark matter (Dey et al., 1998) 
has been used for constructing
numerical models of static
strange stars and strange stars rotating with two different fixed
frequencies 200 Hz and 580 Hz, and for
calculating frequencies of marginally stable orbits around these stars
 (Datta et al., 2000). The authors
conclude that very high QPO frequencies in the range of $1.9$ to $3.1$\,
kHz would imply the existence of a non-magnetized strange star in an X-ray
binary rather than a neutron star.  Gondek-Rosi{\'n}ska
et al. (2001) calculated the maximum orbital frequencies for the Dey
et al. quark stars rotating at various rotation rates and showed that
the rotational and maximum orbital frequencies for these
stars are indeed much higher than those for neutron star models and for
strange stars in the MIT bag model. The {\sl maximum} orbital frequencies
for the compact objects modeled with
 the Dey et al. (1998) equation of state are
$\ge 1.5\,$ kHz, and always higher than the kHz QPO frequencies
observed to date.

Recently, new complications in the problem were
pointed out. The innermost stable circular orbit exists also in
Newtonian gravity (Klu{\'z}niak,
Bulik, Gondek-Rosi{\'n}ska 2000; Amsterdamski et al., 2000; Zdunik \&
Gourgoulhon 2001).
The presence of a crust has been investigated for normal
evolutionary sequences by Zdunik, Haensel \& Gourgoulhon (2001).
Typically the crust increases the maximum orbital frequency 
at the Keplerian limit. 

In this paper we consider constant baryon mass sequences of strange
stars described by the simplest MIT bag model (with massless and
non-interacting particles) for various values of the baryon mass and
calculate the frequency in the co-rotating ISCO (if it exists)
 for each sequence.  We
find and discuss  lower limits to the maximum orbital frequency
(taking into
account various secular and dynamical stability limits, as well as
physical constrains relevant for accretion in LMXBs).
We then extend our results to all MIT
equations of state. Our calculations are described in section 2 and
the results are presented in section 3.  Finally, section 4 contains a
summary of our results.

\section{The model star calculations} 

\begin{figure*}
\centering
\includegraphics[width=17cm]{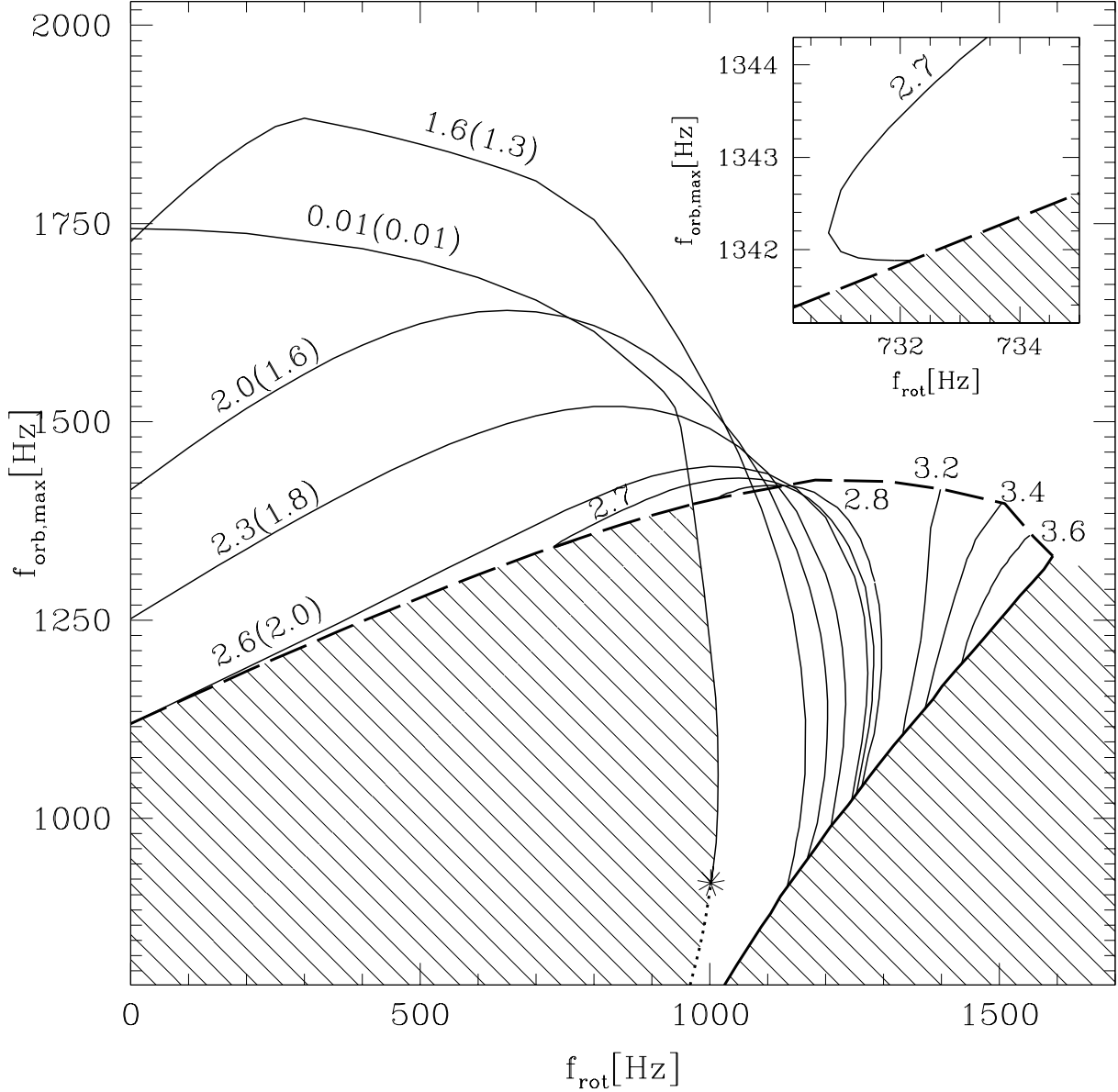}
\caption{The dependence of the maximum orbital frequency on the
frequency of rotation $\Omega/2\pi$.
 Each thin solid line corresponds to a sequence of equilibrium models with
a constant baryon mass, and
is labeled by this baryon mass in solar mass units, as well as
(in parentheses) by the gravitational mass of the static configuration, if
it exists. The dashed line indicates  stars
marginally stable to axisymmetric perturbations.
The angular momentum increases along each sequence from $J=0$
for static configurations, or $J_ {\rm min}$ on the dashed line
for supramassive stars, to
$J_{\rm max}$ for the Keplerian configurations (thick solid line).
  One sequence for a very low mass star (with 
$M_{\rm b}=0.01 M_\odot$) is shown, the critical point  on this sequence
for Newtonian
dynamical instability to non-axisymmetric perturbations is indicated
by an asterisk and dynamically unstable configurations
are denoted with the dotted line.The hatched area
corresponds to a region  excluded for rotating strange stars in the
model considered. The inset shows in detail how the supramassive
sequence reaches the stability limit to quasi-radial
oscillation.}
\end{figure*}

Typically, strange stars are modeled
(Alcock et al., 1986; Haensel et al., 1986)
with an equation of state based
on the MIT-bag model of quark matter, in which
quark confinement is described by an energy term proportional to the
volume (Farhi and Jaffe 1984).
 The equation of state is given by the simple formula
$$P=a(\rho-\rho_0)c^2, \eqno (1)$$
 where $P$ is the pressure, $\rho$
the mass-energy density, and $c$ the speed of light.
It was shown by Zdunik
(2000) that all MIT bag models can be approximated very well by
equation (1), where $a$ is a constant of model-dependent value in the
range 0.289 (for a strange quark mass of ${\rm m_s}$= 250 MeV) to 1/3
(for massless quarks).
 In the numerical calculations reported in the present paper
we describe strange quark matter using
the simplest MIT bag model (with massless and non-interacting
quarks), accordingly we take $a=1/3$ and
$\rho_0=4.2785\times  10^{14}\ {\rm g\ cm}^{-3}\equiv \rho_{\rm fid}$
 (corresponding to a bag
constant of $B=60\ {\rm MeV\ fm^{-3}}$). In general,  eq. (1)
corresponds to self-bound matter with  density $\rho_0$ at
zero pressure and with a fixed sound velocity ($\sqrt{a}\,c$)
at all pressures.
For a fixed value of $a$, all
stellar parameters are subject to scaling relations with
appropriate powers of $\rho_0$ (see, e.g., Witten 1984; Zdunik 2000).
 Approximate scalings with $a$ were
considered by Stergioulas et al. (1999) for the limiting
case of rotation at Keplerian frequencies.
 Although a precise scaling-law with $a$ for other cases
has not been derived, we will show that the lower
limits on the maximum ISCO orbital frequency obtained for the
simplest MIT bag model can easily be used to find lower limits on the
maximum orbital frequency for any other value of $a$ and $\rho_0$.

We have calculated the maximum orbital frequency around uniformly
rotating strange stars using two different highly accurate relativistic codes. 
The first code is based on
the multi-domain spectral methods developed by Bonazzola et al.
(1998). This method has been used previously for calculating 
properties of rapidly rotating strange stars described by the MIT bag
model (Gourgoulhon et al., 1999) and by the Dey model (Gondek-Rosi{\'n}ska
et al., 2000a). This code was also used for calculating the maximum
orbital frequency around normal sequences of MIT strange stars (Zdunik et
al., 2000b, Zdunik \& Gourgoulhon 2001) and of the Dey model strange
stars (Gondek-Rosi{\'n}ska et al., 2001). The second code 
(Stergioulas \& Friedman 1995, see Stergioulas 1998 for a description) 
has been used
to compute accurate constant baryon mass sequences and models at the Keplerian
mass-shedding limit, 
where the first code has difficulties converging to a solution. 
In this code the equilibrium models are obtained following the KEH method
 (Komatsu et al., 1989), in which the field equations are converted to
integral equations using appropriate Green's functions. Models computed by both
codes agree very well in all computed properties.

We calculate the maximum frequency of a co-rotating test particle in a stable
circular orbit in the equatorial plane about a strange star,
 $f_{\rm orb, max}$.
By testing stability of the orbital motion, we determine whether
or not stable orbits extend to the surface of the star (at radius $R_*$).
If they do, the maximum orbital frequency is the keplerian frequency at
the surface, $f_{\rm orb,max}= f_{\rm orb}(R_*)$.
If they do not, we find the radius of
the innermost marginally stable orbit, $R_{\rm ms}$, and its frequency
$f_{\rm ms}$ (see, e.g. Friedman, Ipser \& Parker 1986). 
In this latter case,
the orbital frequency in the ISCO is the maximum circular orbital frequency,
$f_{\rm orb,max}=f_{\rm ms}$ . 
 This frequency can then be compared with
the observed kHz QPOs, having in mind some model of QPOs relating
their frequency to the orbital frequency and the location of the ISCO
(e.g., Klu\'zniak et al., 1990,
Abramowicz and Klu\'zniak 2001, Wagoner et al., 2001).
 In so doing, we neglect the effects of fluid pressure,
magnetic fields,  radiation drag, etc.,
on the location of the innermost circular orbit in the accretion disk.

We construct equilibrium sequences of rotating compact strange stars
with constant baryon mass. We compute both normal and supramassive
sequences.  A
sequence is called normal if it terminates at the zero angular
momentum limit with a static, spherically symmetric solution, and it
is called supramassive if it terminates at the axisymmetric stability limit,
instead.  The boundary
between these two sequences is the sequence with the maximum baryon
mass of a static configuration.  The angular momentum (and the central
density) of a star changes monotonically along each sequence.

For a supramassive constant baryon mass sequence a model is stable to
axisymmetric perturbations, if $\displaystyle \left( {\partial J \over
\partial \rho_{\rm c}} \right)_{M_{\rm b}} < 0$ (or $\displaystyle
\left( {\partial M \over \partial \rho_{\rm c}} \right)_{M_{\rm b}} <
0$) where $J$ is the angular momentum, $M$ is the gravitational mass,
$M_{\rm b}$ is the baryon mass of the star and $\rho_{\rm c}$ is the
central density of the star (Friedman, Ipser \& Sorkin 1988; Cook,
Shapiro \& Teukolsky 1994).

The star in a sequence reaches the Keplerian limit (the mass-shedding
limit) if the velocity at the equator of a rotating star is equal to
the velocity of an orbiting particle at the surface of the star.

\section{Results}

 The properties of the ISCO around strange stars differ from those
around neutron stars (Miller, Lamb \& Cook 1998)---for 
strange (i.e., quark) stars of any mass
the ISCO always exists at sufficiently rapid stellar rotation rate
(Stergioulas et al., 1999; Zdunik et al., 2000b), and for strange stars of 
moderate and high baryon masses the ISCO always exists for any
rotation rate (Zdunik et al., 2000b; Gondek-Rosi{\'n}ska et al., 2001).
For lower mass stars (e.g., the $1.6M_\odot$ baryon mass sequence in
Fig. 1) 
 stable orbits extend down to the stellar surface
for moderate rotation rates, but the gap is present for both slowly
and rapidly rotating stars.  For lower masses yet,  the gap is
present only at high rotation rates and the ISCO is present even if
relativistic effects are negligible (as in substellar-mass configurations,
e.g., for $M\le0.01M_\odot$). In the Newtonian limit, the gap
between the marginally stable orbit and the stellar surface is
produced by the oblateness of the rapidly rotating low-mass quark star
(Klu\'zniak, Bulik \& Gondek-Rosi\'nska, 2000; Amsterdamski et al., 2000;
 Zdunik \& Gourgoulhon 2001).
For Keplerian configurations the ISCO is
always well above the stellar surface for all possible masses.

\begin{figure}
\includegraphics[width=0.9\columnwidth]{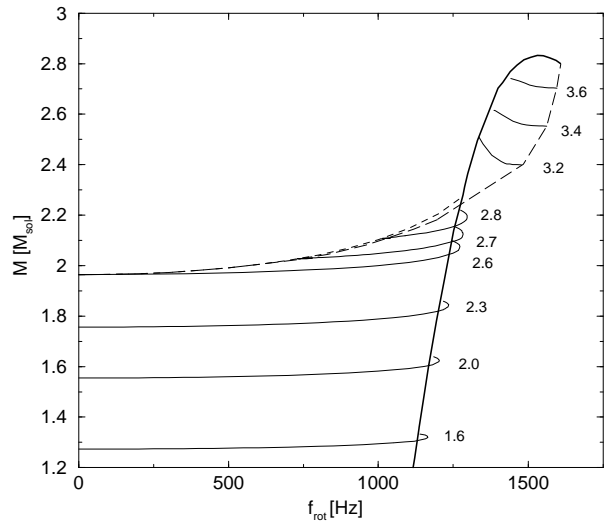}
\caption{Gravitational mass as a function of the rotational frequency
  for the same constant baryon mass sequences as shown in Figure~1. The
  symbols are the same as in Fig~1. The short dashed line corresponds to the
  upper limit on gravitational mass for the model considered.}
\label{wrys2}
\end{figure}

We present the dependence of the maximum orbital frequency on the
frequency of rotation $f_{\rm rot}=\Omega/2\pi$ for a set of
constant baryon mass sequences of strange stars in Figure~1. In
Figure~2 we show how the gravitational mass depends on rotational
frequency for the considered sequences. The normal sequences start at
$f_{\rm rot}=0$, and continue with increasing angular momentum up to
the Keplerian configuration. As it turns out,
the very low mass limit, represented in
Figure~1 by the case of $M_{\rm b}=0.01 M_\odot$, is described well by
a uniform-density fluid, i.e., the Maclaurin spheroid (Chandrasekhar
1969). For this sequence, an ISCO exists only for rotation rates
larger than $\sim 970 \ {\rm Hz} $, and we indicate
the onset of the (Newtonian)
dynamical instability to non-axisymmetric perturbations
by an asterisk.

The intermediate-mass case is shown by the constant baryon mass
sequence with the baryon mass $M_{\rm b} = 1.6 M_\odot$, for which
the ISCO does not
exist between the rotation rates with frequency of $f_{\rm rot}
\approx 300$\,Hz and $f_{\rm rot}\approx 800$ Hz (see also Zdunik et
al., 2000b).

More massive stars in normal sequences are shown for the cases of
$M_{\rm b}=2.0 M_\odot$ and $2.3 M_\odot$. In these cases, the
marginally stable orbit exists along the entire constant baryon mass
sequence. The ISCO frequency initially increases with rotational
frequency, as calculated in the slow-rotation approximation by
Klu{\'z}niak \& Wagoner 1985. When the small-rotation-rate
assumption is no longer satisfied, the ISCO frequency reaches a
maximum and begins to decrease mainly due to the rotational
deformation of the star (Sibgatulin \& Sunyaev 1998; Shibata \& Sasaki 1999).
 The sequence terminates when the star has
reached the Keplerian configuration (mass-shedding limit). An extreme case
of normal sequences is the sequence starting with the maximum mass
static configuration of $M_{\rm b}=2.624 M_\odot$ (indicated as 2.6
$M_\odot$ in Figures~1 and 2).

Supramassive sequences start at the onset of the instability to
axisymmetric perturbations (thick dashed line in Figure~1
and in subsequent
figures.)
Sequences with $M_{\rm b}=2.7 M_\odot$ and $2.8 M_\odot$ are examples
of low mass supramassive sequences. The behavior of the maximum
orbital frequency vs. frequency of rotation is basically similar to the
normal sequences described above.

Finally, we present the sequences of models with $M_{\rm b}=3.2,\,
3.4,$ and $3.6 M_\odot$. For these models of extremely high mass,
 the rotational frequency
decreases with increasing angular momentum because the stars are
initially highly deformed. The ISCO frequency only decreases along
these sequences until the stars reach the mass shedding limit.
The configuration with the highest rotation frequency
is not Keplerian. For a high-mass supramassive
strange star sequence (e.g., $M_{\rm b}=3.2$), the Keplerian
configuration is the one with the lowest rotational frequency in the
sequence (see Fig.~2).   

Note
that the rotational frequency does not necessarily change
monotonically along a constant baryon mass sequence, since a star
changes its shape and, consequently, its moment of inertia with
increasing angular momentum (see also Cook, Shapiro \& Teukolsky 1992;
Zdunik et al., 2000b; Gondek-Rosi\'nska et al., 2000a). 
For a lower-mass supramassive sequence and a
normal sequence the rotational frequency decreases with increasing
angular momentum just below the mass-shedding limit (characteristic
turning back in Figure~1 and 2 for each sequence). A similar behavior
is observed in the case of models with small mass, e.g. for the
constant baryon mass sequence with $M_{\rm b}=0.01 M_\odot$. Thus, the
effect of spin-down with increasing angular momentum is of Newtonian
origin, and is related to the change of shape of the star.

In Figure~1 one can see that the value of the ISCO frequency alone is
not a sufficient indicator for the mass of the compact object. For a
rapidly rotating strange star the same value of orbital frequency at
the innermost stable circular orbit, e.g. 1.2 kHz, can correspond to a
range of stars differing in mass by a factor of a hundred, or even
more, as stressed already by Klu\'zniak, Bulik \& Gondek-Rosi\'nska
2000.  However, it is worth noting that for {\sl rotational}
frequencies exceeding about 1.3 kHz, i.e., for rapidly rotating
supramassive stars, the rotational period alone constrains the stellar
mass to a narrow range.

\subsection{Lower limits on the ISCO frequency}

\begin{figure}
\includegraphics[width=0.9\columnwidth]{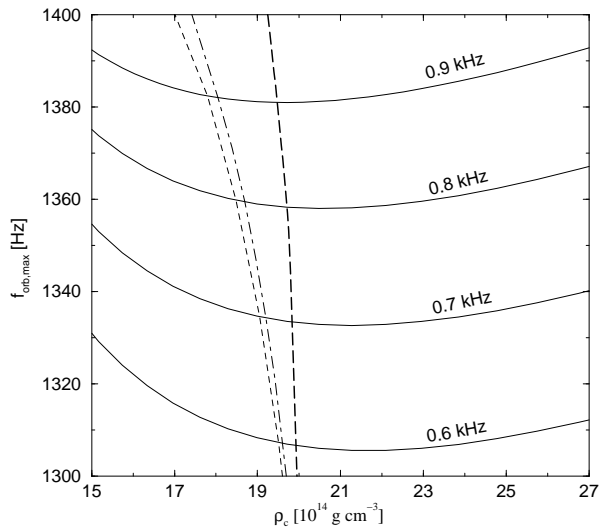}
\caption{The ISCO frequency as a function of central density for
sequences with a fixed rotational frequency, as labeled.
The short dashed line denotes the configurations with the
maximum possible gravitational mass for given frequency (see Figure
2), the dash-dotted line corresponds to stars with maximum baryon
mass, and the thick dashed line represents the axisymmetric stability
limit.  }
\label{marg}
\end{figure}

Figure~1 shows that at any given stellar rotational frequency
the maximum orbital
frequency $f_{\rm orb,max}(f)$ is limited from below by the least
frequency  among the following three ``extreme''
sequences of stars: a) the marginally stable configuration to
axisymmetric perturbations (thick dashed line); b) configurations
rotating at Keplerian frequencies (thick solid line) and c) the
Newtonian limit of low-mass stars, tending to Maclaurin spheroids
 when $M\to 0$, $J \to 0$, $\rho \to\rho_0$
(solid line ending with the asterisk).  For all these
limiting cases, an ISCO exists.
The maximum orbital frequencies are also bounded from {\sl above}
by a fourth limiting line (not shown), which will be discussed elsewhere:
the envelope of highest values of $f_{\rm orb,max}$.

a) We find that at a given low or moderate value of stellar rotational
 frequency the
lowest ISCO frequency is obtained for the configurations at the
axisymmetric instability limit.  As was shown by Gourgoulhon et al.
(1999) and Gondek-Rosi{\'n}ska et al. (2000a) a supramassive strange star,
just as a supramassive neutron star (Cook et al., 1992),
prior to its collapse to a
black hole increases its spin as it loses angular momentum. In such
cases, two models with the same baryon mass but with different central
density and angular momentum can have the same rotational frequency
(see inset in Figure~1).  Along a sequence of fixed rotational
frequency, the model with the lowest ISCO frequency is not the model
with the highest gravitational or highest baryon mass.  The dependence
of the ISCO frequency as function of the stellar central density for
sequences of stars with a fixed rotational frequency is shown in
Figure~\ref{marg}.  As the central density increases, the ISCO
frequency decreases and the sequence passes through the configuration
with maximum gravitational mass (denoted with short dashed line in
Figure~\ref{marg} and Figure 2), and then the maximum baryon mass
configuration (denoted with the dash-dotted line). Before reaching the
minimal value of $f_{\rm orb,max}$,
 the sequence reaches the limit of stability to
axisymmetric perturbations (thick dashed line). Thus, for a fixed
rotational frequency, the lowest possible maximum orbital frequency is
attained for the model which is marginally stable to gravitational
collapse.

b) At high rotation rates the lowest ISCO frequency is attained for the
 Keplerian (mass-shedding) configurations (thick solid line in
 Figure~1). The stellar configurations at this limit are highly
 deformed.  We calculated Keplerian configurations beginning
 with the fastest rotating stable configurations (with $M_{\rm
 b}=3.76 M_\odot$), ending with a strange star
 with $M_{\rm b}=0.4 M_\odot$. Note that for rapidly
 rotating (close to the Keplerian limit) strange stars, the ISCO
 frequency is smaller than the rotational frequency.

c) In the small-mass limit  the ISCO only
exists at high stellar rotation rates. This was discovered numerically,
and the Newtonian origin of the ISCO in this case
was first understood by comparison with
toy models of disk-like distributions of matter
(Klu{\'z}niak, W., Bulik, T.,\& Gondek-Rosi{\'n}ska 2000;
Zdunik \& Gourgoulhon 2001). In the Newtonian limit 
($M\to 0$, $J/M^2\to\infty $) strange stars
tend to a nearly uniform-density configuration
($\rho \to\rho_0$), and these
can be described well by the Maclaurin spheroids,
for which analytic formulae for orbital frequencies have been
derived and found to agree with the numerical models
(Amsterdamski et al., 2000).
The functional dependence of the
maximum orbital frequency on the stellar rotation rate 
is nearly universal for all strange stars with masses lower
than 0.1 $M_\odot$, in the model considered here.

\subsection{The lower limits on ISCO for other MIT bag models}

\begin{figure}

\includegraphics[width=0.9\columnwidth]{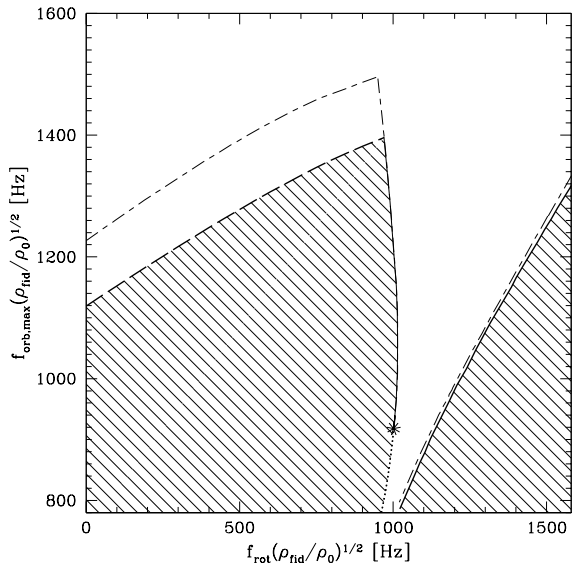}
\caption{The effect of non-zero strange quark mass on the lower limits
on the maximum orbital frequency. The dependence on
$\rho_0$ has been scaled out.
Here, the thin dotted-dashed lines correspond to the
limits for the case of ${\rm m_s}=250\, $MeV (a=0.289),
 and all  other lines and symbols have the same meaning as in Figure~1,
which was obtained for $a=1/3$ ( ${\rm m_s}=0$). 
 }
\label{scale}
\end{figure}

Figure~1 and the discussion above assumed the simplified MIT bag model
with massless and non-interacting quarks.  The results can easily be
scaled for any value of the density at zero pressure, with the use of
the scaling $f \propto \rho_0^{1/2}$, where $f$ denotes either of the
frequencies (rotational or orbital), and $M \propto \rho_0^{-1/2}$.  We
can easily find lower limits on ISCO for any other MIT bag model since
Zdunik (2000) showed that all MIT bag model equations of state can be
well approximated by eq. (1). 

In order to find the dependence of the lower limits on $a$ we display
in Figure~\ref{scale} the limits calculated for ${\rm m_s}=0$ and for
${\rm m_s}=250\,$MeV.  We have chosen these values of the strange
quark mass because it corresponds to the highest value $a=0.333$, and
the lowest value of $a=0.289$ for the MIT bag model equation of state.
The dependence on the density of quark matter at zero pressure has
been scaled out by multiplying the frequencies with the square root of
$\rho_{\rm fid}/\rho_{0}$, where our fiducial value of $\rho_0$ was
used: $ \rho_{\rm fid}=4.2785\times 10^{14}\ {\rm g\ cm}^{-3}$.  The
numerical values of the frequencies in Fig. 4 correspond to the value
of the bag constant $B=60\ {\rm MeV\ fm^{-3}} $ used throughout this
paper.

 The limit on the
maximum orbital frequency, set by the low mass stars does not depend
on the value of $a$. The limit at rapid rotation rates depends only
slightly on the value of $a$. The main difference appears in the slow
and moderate rotation regions, where the lower limit is set by the
sequence of marginally stable stars.  However, here one can use the
fact that the limit is only shifted along the vertical axis. 
We find that the formula
$f_{\rm orb,max}(a,f_{\rm rot})= 0.3~f_{\rm rot} +f_{\rm orb,max}(a,0)$ is a
good approximation for describing the lower limit on the ISCO frequency
for slowly and moderately
rotating strange stars described by any MIT equation of state.
 Here, the maximum
orbital frequency of the nonrotating model is 
given by its Schwarzschild-metric value
$$f_{\rm orb,max}(a,0)=2.2\,{\rm kHz} /(M(a)/{\rm M_\odot})$$ 
(e.g., Klu\'zniak et al., 1990), and $M(a)$ is the maximum mass in general
relativity of a static configuration of the (quark) fluid described by
 eq. (1)---in short, the maximum mass of a non-rotating strange star.

\subsection{Factors that could alter the lower limits}

\begin{figure}
\includegraphics[width=0.9\columnwidth]{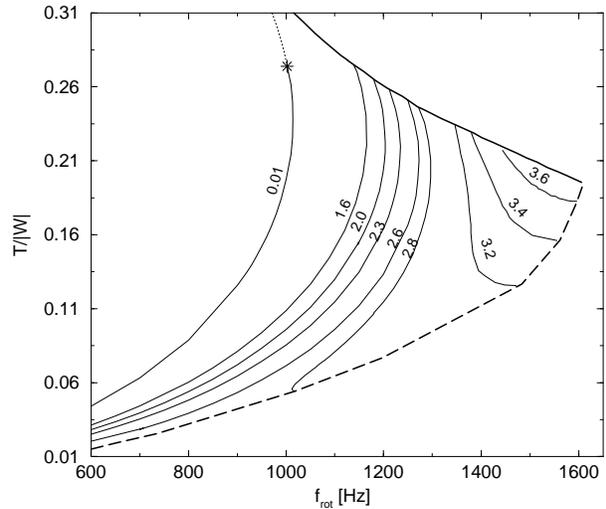}
\caption{ The ratio of the kinetic energy to the absolute value of the
  gravitational potential energy (defined as in Friedman et al., 1986)
  as a function of the rotational frequency for several constant baryon mass
  sequences of MIT bag model strange stars.  The meaning of lines is the same
  as in Fig 1. The critical value of $T/|W|$ for the onset of dynamical
  stability in the low mass limit is the Newtonian value of $0.27$
  indicated by an asterisk.}
\label{tw}
\end{figure}

There are several factors which could affect the lower limits
considered in the previous section. We discuss here: i) different kinds of
instabilities that could operate in rapidly rotating strange stars,
ii) the existence of a solid crust on the accreting strange stars,
iii) details of the accretion-induced spin-up of strange stars in LMXBs.

i) Rapidly rotating compact stars may be subject to non-axisymmetric
rotational instabilities. A widely used indicator for the onset of
instability is the ratio of the kinetic energy to the absolute value
of the gravitational potential energy $T/|W|$.  We plot the value of
$T/|W|$ (defined as in Friedman et al., 1986) for several
constant baryon mass sequences in Figure~\ref{tw}.  The values of
$T/|W|$ for strange stars rotating at the Keplerian velocity are very
large and can be even higher than $T/|W|=0.32$.\footnote{In
Gourgoulhon et. al., 1999 (see their Fig.~7), $T/|W|<0.24$, because
configurations denoted as Keplerian, were in fact configurations at
which the numerical procedure stopped converging.}  The $T/|W|$ is
very high for all linear self-bound equations of state (see
Gondek-Rosi\'nska et al., 2000b for different MIT bag models
and Gondek-Rosi\'nska et al., 2000a for the strange stars described by
the Dey model). The values of $T/|W|$ for strange stars are much
larger than for any neutron star models (for which the maximum $T/|W|$
ranges typically from 0.1 to 0.14, (see e.g.  Nozawa et al., 1998). In
contrast to neutron stars, the $T/|W|$ for strange stars increases
with decreasing stellar mass for all models.  Very large values of
$T/|W|$ are characteristic not only for supramassive stars but also
for rapidly rotating strange stars with small and moderate masses. The
large value of $T/|W|$ results from a flat density profile combined
with strong equatorial flattening of rapidly rotating strange stars.
No stability analysis in general relativity has been performed yet for
strange stars.

There exists two different classes and corresponding time-scales for
non-axisymmetric instabilities: a dynamical instability, growing on
the hydrodynamical time-scale and secular instabilities (due to
dissipative mechanisms) which grow on much longer time-scales.
In Newtonian theory, rapidly rotating
Maclaurin spheroids are secularly unstable to a bar-mode formation, if
$T/|W| > (T/|W|)_{\rm crit} = 0.1375$ (Jacobi/Dedekind bifurcation point)
and dynamically unstable if ($T/|W| > 0.2738$).
A secular instability can grow only in the presence of dissipative
mechanisms, such as shear viscosity (Roberts \& Stewartson 1963) or
gravitational radiation (the Chandrasekhar-Friedman-Schutz instability
(hereafter CFS instability; Chandrasekhar 1970; Friedman \& Schutz
1978; Friedman 1978).  If both viscosity and gravitational radiation
are operative, they act against each other.  Which of the two dissipative
mechanisms dominates is highly dependent on the temperature of the
star.  By a suitable choice of the strength of viscosity relative to
gravitational radiation, it is possible to stabilize the Maclaurin
sequence up to the point of dynamical instability (Shapiro \&
Teukolsky 1983).

In general, the viscosity-driven instability could be operating in
accreting (old and cold) compact objects, while the CFS instability
could operate in both newly-born compact stars and in LMXBs. The most
important pulsation modes for the CFS instability are the $r$- and
$f$-modes (see Stergioulas 1998, Andersson \& Kokkotas 2001, Friedman
and Lockitch 2001, for recent reviews of these instabilities).

The dynamical instability against bar-mode deformation of rapidly
rotating polytropic stars was studied by Shibata, Baumgarte
\& Shapiro (2000) in full general relativity and by Saijo et al., 2001
in the first post-Newtonian approximation. The authors considered
differentionally rotating proto-neutron stars. Their
main result is that in general relativity the onset of dynamical
instability occurs for a somewhat smaller critical value of $(T/|W|)_{\rm
  crit}\sim 0.24-0.25$ than the Newtonian value $\sim 0.27$ for
incompressible Maclaurin spheroids.

The few numerical studies of secular instabilities in rapidly rotating
 stars, in general relativity show that nonaxisymmetric modes driven
 unstable by viscosity no longer coincide with those driven unstable
 by gravitational radiation. The $(T/|W|)_{crit}$ depends not only on
 the dissipative mechanism but also on the compaction parameter $M/R$
 of a star.  The viscosity-driven instability for neutron stars has
 been studied in general relativity by Bonazzola, Frieben \&
 Gourgoulhon (1996, 1998), who find that it sets in for $T/|W|$
 somewhat larger than the Newtonian value of $(T/|W|)_{\rm crit}=
 0.1375$. A stronger weakening of the viscosity-driven instability by
 general relativity is suggested by a post-Newtonian study of the
 viscosity-driven instability in incompressible stars by Shapiro \&
 Zane (1998), who find that the $(T/|W|)_{\rm crit}$ could be as large
 as 0.25 for $M/R\sim 0.2.$ On the other hand, the
 gravitational-radiation driven instability is strengthened by
 relativity and the bar ($l=2$) $f$-mode can become unstable for
 typical neutron star equations of state (Stergioulas \& Friedman
 1998). For a wide range of realistic equations of state, Morsink,
 Stergioulas \& Blattning (1999) find that the critical $T/|W|$ for
 the onset of the instability in the $l=2$ $f$-mode is only $T/|W|
 \sim 0.08$ for models with mass $1.4 M_\odot$ and $T/|W| \sim 0.06$
 for maximum mass models.

The CFS-instability of $r$-modes in strange stars has been studied by
Madsen (2000), in the Newtonian limit. According to this study, the
instability can operate at a minimum rotational frequency of $\sim
400$Hz and at a temperature of $T\sim 10^7$K. Since strange stars can,
at most, have only a thin solid crust, they cannot store heat in the
interior during accretion.  Thus, accreting strange stars are likely
to have interior temperatures comparable to $T\sim 10^7$K. If the
$r$-mode instability is not damped by other mechanisms then the above
results seem to rule out the presence of low and moderate mass stars
with a normal quark matter in LMXBs (if the observed 1.07 kHz QPO is
the orbital frequency at the ISCO).

If, contrary to our current understanding, the $r$-mode instability is
not effective in accreting strange stars, then the $f$-mode
instability or the viscosity-driven instability could be more
relevant. More detailed studies of these instabilities specifically
for relativistic strange stars are needed. The dynamical instability,
even if present at $T/|W|\sim 0.24$, only marginally affects the
limits presented in Fig.~1.

ii) Another factor affecting the lower limits presented in Figure~1
would be the presence of a solid crust. This effect has been discussed
by Zdunik, Haensel \& Gourgoulhon (2001).  The presence of the crust
will not affect the lower limit in the small and moderate rotation
case.  Here the stability of the strange stars is basically determined by the
stability of the strange core and the properties of the crust are
negligible. However, for rapid rotation, when the lower limit on the
maximum orbital frequency is determined by the Keplerian
configurations, the presence of the crust will increase the lower
limit.  The crust is of low density and for rapidly rotating strange stars it
becomes extended in the equatorial region. Thus the mass-shedding
limit is reached at lower rotation rates than in the case of bare
strange stars.  Zdunik, Haensel \& Gourgoulhon (2001) showed that the
values of maximum orbital frequency for strange stars with maximum
crust (up to the neutron drip point), rotating at the mass-shedding
limit, are about $20\%$ larger compared to case of bare strange stars.  Also,
for strange stars  with a crust,
 the ISCO exists only for models with large mass.

iii) When a strange star is spun-up by accretion of material that
leaves the ISCO and plunges onto the surface of the star, the largest
attainable rotation rate is not precisely the rotation rate at the
mass-shedding limit. Material that leaves from the ISCO arrives at the
surface of a strange star after following a path in which it conserves
angular momentum. At the surface, the material has angular velocity
larger than what it had at the ISCO, but still somewhat smaller than
the angular velocity needed to spin-up the star to the angular
velocity of the equilibrium configuration at the mass-shedding
limit. The difference between the maximum rotation and mass-shedding
configurations is, however, small.

\section{Summary}

Relativistic orbital frequencies are crucially related to the spin-up
of neutron stars and black holes and to high energy accretion
phenomena in low mass X-ray binaries, including quasi-periodic
oscillations in the X-ray flux.  We compute the maximum orbital
frequency of stable circular motion around strange stars, and present
it as a function of the stellar rotation rate for normal and
supramassive constant baryon mass sequences of stars rotating at all possible
rates. We find that in contrast to neutron stars, the mass of
a  rapidly rotating strange star cannot be even approximately
inferred from the orbital frequencies alone. For
strange stars described by the MIT bag model, the same
frequency in the innermost stable orbit (e.g., 1.2 kHz) is obtained
for strange stars with rotational periods ranging from infinity to
about 0.6 ms, and the mass ranging from that of a planetoid to about
three solar masses.  One should note however that the
rapidly rotating stars in this region are quite likely to be unstable.

 We find that for slow to moderate rotation rates
the lower limit on the maximum orbital frequency around
strange stars is attained for configurations marginally stable to
axisymmetric perturbations,
while at rapid rotation rates it is reached 
for configurations at the mass-shedding limit.
There are several factors which could affect the lower limit for
rapidly rotating strange stars, such as different instabilities or the
existence of a solid crust. The $r$-mode instability has the potential
to rule out strange stars with small and moderate masses in LMXBs
(assuming that the observed 1.07 kHz QPO in 4U 1820-30 is close to
the Keplerian frequency at the innermost stable orbit of an accretion
disk around a strange star). If the $r$-mode instability is not
effective, then the $f$-mode instability and the viscosity-driven
instability may become relevant.  On the other hand, the dynamical
instability only marginally affects our computed lower limits on the
maximum orbital frequency. Our results apply to all MIT bag models.

\begin{acknowledgements}
 This work has been funded by the following grants:
KBN grants 5P03D01721, 2P03D00418, 2P03D01816; the Greek-Polish Joint
Research and Technology Programme EPAN-M.43/2013555 and the EU
Programme "Improving the Human Research Potential and the
Socio-Economic Knowledge Base" (Research Training Network Contract
HPRN-CT-2000-00137).
\end{acknowledgements}


\begin{thebibliography}{} 
\bibitem[]{}
Abramowicz M., Klu{\'z}niak W., 2001, A\&A, 374, L19

\bibitem[]{}
Alcock C., Farhi E., Olinto A., 1986, ApJ 310, 261

\bibitem[]{}
Amsterdamski P., Bulik T., Gondek-Rosi\'nska D., Klu\'zniak W., 2000,
astro-ph/0012547

\bibitem[]{}
Andersson, N., Kokkotas K. D., 2001, Int. J. Mod. Phys. D, in press,
 gr-qc/0010102 

\bibitem[]{}
Bombaci I., Datta B., 2000, ApJ, 530, L69

\bibitem[]{}
Bonazzola S., Frieben J., Gourgoulhon E., 1996, ApJ, 460, 379

\bibitem[]{}
Bonazzola S., Frieben J., Gourgoulhon E., 1998b, A\&A, 331, 280

\bibitem[]{}
Bonazzola S., Gourgoulhon E., Marck J. A. 1998a, Phys. Rev. D., 58, 104020

\bibitem[]{}
Bulik T., Gondek-Rosi{\'n}ska D., Klu{\'z}niak W., 1999a, A\&A, 344, L71

\bibitem[]{} Bulik T., Gondek-Rosi{\'n}ska D., Klu{\'z}niak W.,
1999b, ApJ Letters and Comunications, 38, 77

\bibitem[]{}
Chandrasekhar S., 1969, {\it Elipsoidal figures of equilibrium} (Yale
University Press, New Haven) 

\bibitem[]{}
Chandrasekhar S., 1970, Phys. Rev. Lett., 24, 611

\bibitem[]{}
Cook G. B., Shapiro S. L., Teukolsky S. A., 1992, ApJ, 398, 203

\bibitem[]{} 
{Cook} G.~B., {Shapiro} S.~L., {Teukolsky} S.~A., 1994, ApJ {424}, 823

\bibitem[]{}
Datta B., {Thampan} A.~V., {Bombaci}, I., 2000,
 {355}, L19

\bibitem[]{}
Dey M., {Bombaci} I., {Dey} J., {Ray} S., {Samanta} B.~C., 1998,
Physics Letters B, {438}, 123

\bibitem[]{}
Farhi E. and Jaffe R.L., 1984, Phys. Rev. D 30, 2379

\bibitem[]{}
Friedman J. L., 1978, Commun. Math. Phys., 62, 247

\bibitem[]{}
Friedman J. L., Ipser J. R., Parker L. R., 1986, ApJ, 304, 115; errata 
published in ApJ 351, 705 (1990)

\bibitem[]{}
Friedman J. L., Ipser J. R., Sorkin R. D., 1988, ApJ, 325, 722

\bibitem[]{}
Friedman J. L., Lockitch K. H., 2001, gr-qc/0102114

\bibitem[]{}
Friedman J. L., Shutz B. F., 1978, ApJ, 222, 281

\bibitem[]{}
Gondek-Rosi\'nska D., Bulik T., Klu\'zniak W., Zdunik J. L.,
    Gourgoulhon E., 2001, Proceedings of the 4th Integral Meeting
    Alicante, in press, astro-ph/0012540

\bibitem[]{}
Gondek-Rosi{\'n}ska D., Bulik T., Zdunik J. L., Gourgoulhon E., Ray S.,
 Dey J., Dey M., 2000a, A\&A, 363, 1005

\bibitem[]{}
Gondek-Rosi{\'n}ska D., Haensel P., Zdunik J. L., {Gourgoulhon} E.,
  2000b, ASP Conf. Ser. 202, 661, astro-ph/0009282

\bibitem[]{}
Gourgoulhon E., {Haensel} P., {Livine} R., {Paluch} E., {Bonazzola} S.,
  {Marck} J.~A., 1999, A\&A, {349}, 851

\bibitem[]{}
Haensel P., Zdunik J.L., Schaeffer R., 1986, A\&A 160, 121 

\bibitem[]{}
Kaaret P., Ford E. C., Chen K. 1997, ApJ, 480, 127

\bibitem[]{}
Kamatsu H., Eriguchi Y., Hachisu I., 1989, Mon. Not. R. Astron. Soc., 237, 
355

\bibitem[]{}
Klu{\'z}niak W., 1998, ApJ, 509, L37

\bibitem[]{}
Klu{\'z}niak W., Bulik T., Gondek-Rosi{\'n}ska D., 2000, 
Proceedings of the 4th Integral Meeting, in press, astro-ph/0011517

\bibitem[]{}
Klu{\'z}niak W., Michelson P., Wagoner R. V., 1990, ApJ, 358, 538

\bibitem[]{}
Klu{\'z}niak W., Wagoner R. V., 1985, ApJ, 297, 548

\bibitem[]{}
Madsen J., 2000, Phys. Rev. Lett., 85, 10

\bibitem[]{}
Miller M. C., Lamb F. K., Cook G.B., 1998, ApJ 509, 793

\bibitem[]{}
Morsink S. M., Stergioulas N., Blattnig S. R., 1999, ApJ, 510, 854

\bibitem[]{}
Nozawa T., Stergioulas N., Gourgoulhon E., Eriguchi Y., 1998, A\&AS, 132, 431

\bibitem[]{}
Roberts P. H., Stewartson K., 1963, ApJ, 137, 777

\bibitem[]{}
Saijo M., Shibata, M., Baumgarte, T. W., Shapiro L., S., 2001, ApJ, 548, 919

\bibitem[]{}
Shapiro S. L, Teukolsky S. A., 1983,{\it Black holes, white dwarfs, and
 neutron stars} (New York: Wiley)

\bibitem[]{}
Shapiro S. L., Zane S., 1998, ApJS 117, 531

\bibitem[]{}
Shibata M., Baumgarte T. W., Shapiro S. L., 2000, ApJ, 542, 453

\bibitem[]{}
Shibata M., Sasaki M., 1999, Phys. Rev. D60 084002

\bibitem[]{}
Sibgatulin N.R., Sunyaev R.A., 1998, Astron. Lett. 24, 774

\bibitem[]{}
Stergioulas N., 1998, Living Rev. Relativity, 1,8,
http:///www.livingreviews.org/Articles/Volume1/1998-8stergio/

\bibitem[]{}
Stergioulas N., Friedman J. L., 1995, ApJ 444, 306

\bibitem[]{}
Stergioulas N., Klu{\'z}niak W., Bulik T., 1999, A\&A, 352, L116

\bibitem[]{}
Wagoner R. V., Silbergleit A. S., Ortega-Rodríguez M., 2001, ApJ,
559, L25

\bibitem[]{}
Witten E., 1984, Phys. Rev. 30, 272

\bibitem[]{}
Zdunik J. L., 2000, A\&A, 359, 311

\bibitem[]{}
Zdunik J.\ L., Bulik T., Klu{\'z}niak W., Haensel P., 
Gondek-Rosi{\'n}ska D.\ 2000a, A\&A, 359, 143 

\bibitem[]{}
Zdunik J. L., Gourgoulhon E., 2001, Phys.Rev. D63,087501

\bibitem[]{}
Zdunik J. L., {Haensel} P., {Gondek-Rosi{\'n}ska} D., {Gourgoulhon} E.,
 2000b, {A\&A} {356}, 612 

\bibitem[]{}
Zdunik J. L., Haensel P., Gourgoulhon E., 2001, A\&A, 372, 535

\bibitem[]{}
Zhang W., Strohmayer T. E., Swank J. H., 1998, ApJ Lett., 482, L167

\end{thebibliography}
\end{document}